\documentclass[a4paper, 10pt, conference]{IEEEtran}
\IEEEoverridecommandlockouts                   

\usepackage{graphicx}
\usepackage{color}
\usepackage{amsmath}
\usepackage{multirow}
\usepackage{booktabs}
\usepackage{url}
\usepackage{svg}
\usepackage{amsmath}
\usepackage{adjustbox}
\usepackage{subfloat}
\usepackage[english]{babel}
\usepackage{comment}
\usepackage{tikz}
\usetikzlibrary{positioning}

\newcommand\copyrighttext{%
  \footnotesize \textcopyright 2020 IEEE. Personal use of this material is permitted. Permission from IEEE must be obtained for all other uses, in any current or future media, including reprinting/republishing this material for advertising or promotional purposes, creating new collective works, for resale or redistribution to servers or lists, or reuse of any copyrighted component of this work in other works. This paper has been accepted for publication in the 2020 Twelfth International Conference on Quality of Multimedia Experience (QoMEX).}
\newcommand\copyrightnotice{%
\begin{tikzpicture}[remember picture,overlay]
\node[anchor=south,yshift=10pt] at (current page.south) {\fbox{\parbox{\dimexpr\textwidth-\fboxsep-\fboxrule\relax}{\copyrighttext}}};
\end{tikzpicture}%
}

\begin{document}

\title{Assessing differences in flow state induced by an adaptive music learning software}

\author{
\IEEEauthorblockN{Martin Haug$^1$, Paavo Camps$^2$, Tobias Umland$^3$, Jan-Niklas Voigt-Antons$^{4,5}$}
\IEEEauthorblockA{$^1$Complex and Distributed IT Systems, TU Berlin, Germany\\$^2$Information Systems Engineering, TU Berlin Germany\\$^3$Computer Science, TU Berlin, Germany\\$^4$Quality and Usability Lab, TU Berlin, Germany\\
 $^5$German Research Center for Artificial Intelligence (DFKI), Berlin, Germany}
}

%\IEEEpubid{\makebox[\columnwidth]{978-1-7281-5965-2-0/20/\$31.00 \copyright 2020 IEEE \hfill} \hspace{\columnsep}\makebox[\columnwidth]{ }}

\maketitle
\copyrightnotice

\begin{abstract}
Technology can facilitate self-learning for academic and leisure activities such as music learning. In general, learning to play an unknown musical song at sight on the electric piano or any other instrument can be quite a chore. In a traditional self-learning setting, the musician only gets feedback in terms of what errors they can hear themselves by comparing what they have played with the score. Research has shown that reaching a flow state creates a more enjoyable experience during activities. This work explores whether principles from flow theory and game design can be applied to make the beginner's musical experience adapted to their need and create higher flow. We created and evaluated a tool oriented around these considerations in a study with 21 participants. We found that provided feedback and difficulty scaling can help to achieve flow and that the effects get more pronounced the more experience with music participants have. In further research, we want to examine the influence of our approach to learning sheet music.
\end{abstract}


%\begin{tikzpicture}[overlay, remember picture]
%\path (current page.north) node (anchor) {};
%\node [below=of anchor] {2020 Twelfth International Conference on Quality of Multimedia Experience (QoMEX)};
%\end{tikzpicture}

%%%%%%%%%%%%%%%%%%%%%%%%%%%%%%%%%%%%%%%%%%%%%%%%%%%%%%%%%%%%%%%%%%%%%%%%%%%%%%%%
\section{INTRODUCTION}

Technology advancements drive the adoption of technical solutions for self-learning \cite{williams2005}. For this as well as for other usage scenarios, it is important that the user can focus on the activity at hand. Sometimes, when performing a task, one can get so immersed that one forgets about almost everything else. This mental state of absolute focus and continuous progress is called flow \cite{Csikszentmihalyi2014}.

In this paper, we investigate whether software and principles of game design can be used to get musicians of any skill level into the flow state while learning to play an unknown song on an electric piano.
Our paper differentiates itself from existing work by considering the inducement of flow during the learning process, rather than for a previously acquired skill. In our study, we also analyze whether prior experience affects the induced flow for the users. We aspire to create a solution that makes reaching the flow state while practicing an unknown song easy, making the learning experience more enjoyable. For this, we created an original audio-visual software tool that can be connected to an electric piano. This software implements difficulty scaling (i.e., adapting pattern complexity) and provides instant visual feedback, two techniques known to facilitate reaching a flow state in disciplines like game design. We perform a repeated measurement study with 21 participants to demonstrate whether these two techniques can help to enter a flow state and to detect significant alterations caused by the musical background of the participants.

In this paper, we would like to investigate whether visual feedback and adaptive difficulty scaling help musicians to achieve flow while playing a new song at sight on the piano. Is the prior research of \cite{Chen2007, Harmat2015} in the domain of games transferable to music? If these measures help musicians to achieve flow, we will investigate the impact of their skill and experience on the strength of the effect. Since flow is considered to be a positive feeling \cite{Csikszentmihalyi2014}, it is a question whether people are interested in a solution that can facilitate flow with these means for practicing their instruments in their own free time. We investigate these hypotheses by asking people with varying musical skill levels about their experience with a self-created tool.

%%%%%%%%%%%%%%%%%%%%%%%%%%%%%%%%%%%%%%%%%%%%%%%%%%%%%%%%%%%%%%%%%%%%%%%%%%%%%%%%
\section{RELATED WORK}
The state and concept of "flow", as initially described by Mihaly Csikszentmihalyi, is based on the repeated encounters of the author with his study participants. He observed that participants of his studies repeatedly described that they were able to enter a state of mind while concentrating, which they likened to being in a water flow that guides them from action to action \cite{Csikszentmihalyi2014}. 
During flow, one focuses all attention on the task at hand. The phenomenon is shown to help achieve peak performance. Workers who report experiencing flow in their work, for example, are recorded to work more productively compared to their peers \cite{Csikszentmihalyi2014}.
Alongside the reported high degrees of creativity and productivity, those who entered the state of flow also preferred returning to their activities, as they associated a positive experience with their task. According to Csikszentmihalyi, a balance between perceived challenges and perceived skill, a clear set of goals as well as immediate constructive feedback help induce a flow state \cite{CsikszentmihalyiEtAl2014}.

Chen suggests that because everyone experiences a stimulus (in his case a computer game) differently, the optimal amount of challenge to experience flow varies from player to player. He, therefore, suggests to let the player pick the difficulty through their actions in the game (dynamic difficulty) \cite{Chen2007}. Harmat et al. showed that increasing the difficulty in a game according to player performance is indeed helpful in achieving flow \cite{Harmat2015}. Rheinberg, Vollmeyer, and Engeser created a survey with ten items to measure the perceived flow of study participants on their "Flow Short Scale" \cite{jackson1996development}. With regards to music and flow, De Manzano et al. used a survey of trained pianists who played self-selected pieces to correlate flow scores to physiological markers \cite{Manzano2010}. Sébastien et al. find seven ways to evaluate the difficulty of a music piece: playing speed, fingering, hand displacement, polyphony, harmony, irregular rhythm, and length \cite{Sebastien2012}. Nakamura et al. created a software that can create piano scores for one player with different difficulties with respect to these criteria using an ensemble/band score \cite{Nakamura2018}.

%%%%%%%%%%%%%%%%%%%%%%%%%%%%%%%%%%%%%%%%%%%%%%%%%%%%%%%%%%%%%%%%%%%%%%%%%%%%%%%%
\section{METHODS}
\subsection{Experimental setup}
\label{sec:tooling}
We have created an application for this experiment that plays a predefined song (left and right hand on the piano). The screen of a TV behind a connected (piano) keyboard displays a stave in which the next notes to play entered from the right side and floated towards a cursor. Once a note hits the cursor, its key should be held down on the keyboard to sync up with the song. A software keyboard is displayed on the bottom section of the screen and highlights which keys should be pressed and are currently pressed. In the adaptive mode, the notes that were played correctly float up to give the user feedback on their performance. Moreover, if the user makes too many mistakes, the software switches to easier patterns. Conversely, if the user is playing everything correctly, the software will switch to more intricate patterns. The base mode has these features switched off. We hypothesize that the changes in the adaptive mode will augment the measured amount of flow.

% rightnote played = -0.75
% wrong note played = 1
% sum( notes in sheets) - sum(notes played
%
%
%
%
% Threeshold for higher difficulty = -20
% Thresshold for lower difficulty = 20

\begin{figure}
    \centering
    \includegraphics[width=0.9\columnwidth]{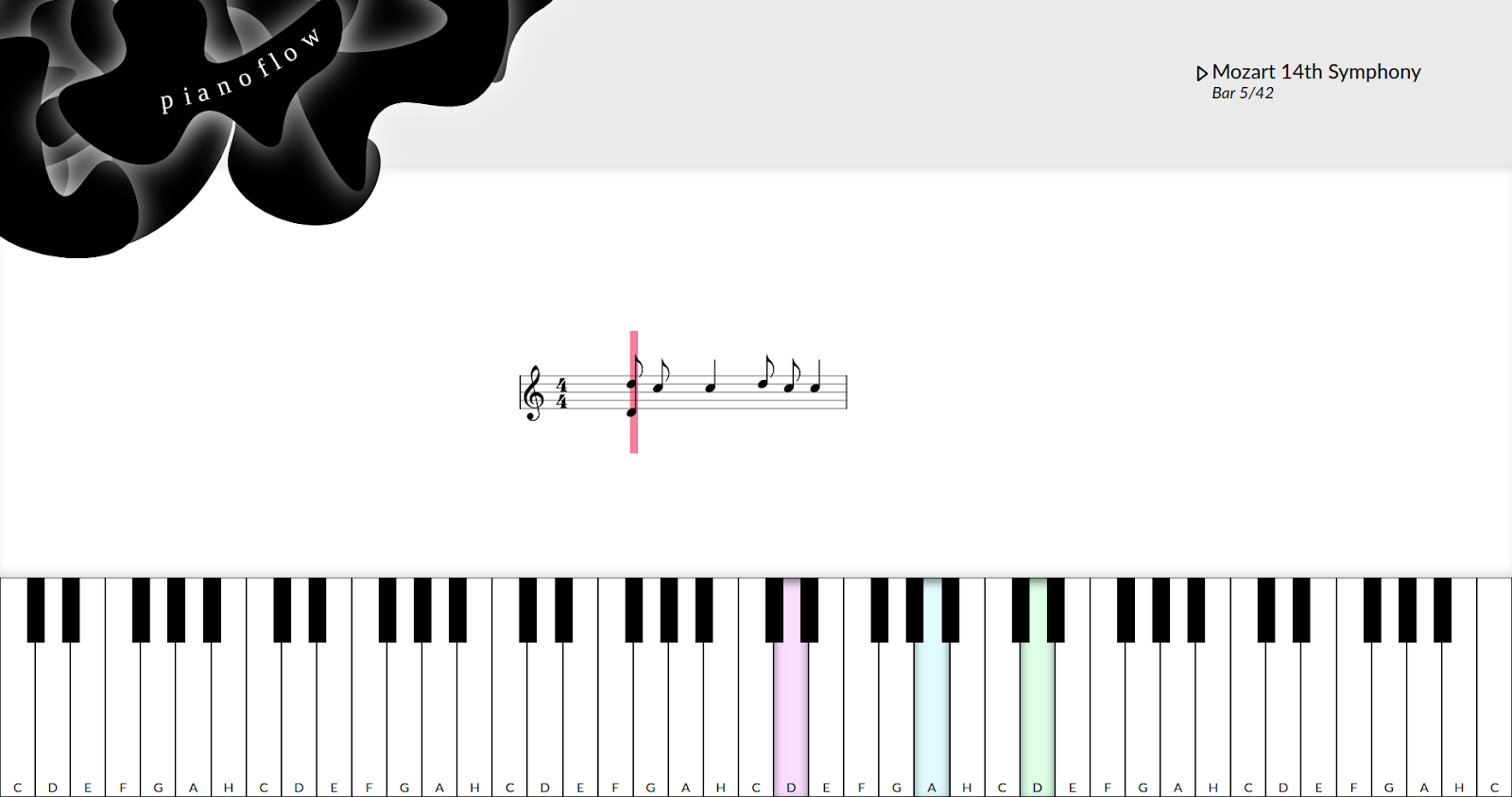}
    \caption{UI for the player while playing a song. The stave holds upcoming notes. The on-screen keyboard highlights held and requested keys and their intersection in different colors.}
    \label{fig:piano-ui}
    \vspace{-2em}
\end{figure}

\subsection{Sampling procedure and sample size}
We intended the test sample size to be at least 20 people. Testing was conducted from 24\textsuperscript{th} to 27\textsuperscript{th} of February 2020 on a set of 21 participants with an average age of 24.7 years. Seven of the participants ($33\%$) identified as female, the remaining 14 participants identified as male. The participants were either invited personally by the authors or were advertised the experiment after having completed other experiments in the same laboratory. There were no special incentives to participate. There were no restrictions on participation. In particular, no prior experience with reading sheet music or musical instruments was required. The study was conducted following ethical principles for medical research involving human subjects proposed by the World Medical Association (WMA) Declaration of Helsinki.

\subsection{Measures and covariates}
For the experiments, we recorded a quantitative measure of flow using the Flow Short Scale \cite{jackson2010flow} (discrete 5-point Likert scale). Another measure was whether a participant preferred the adaptive mode designed to induce flow or the base mode (see Section \ref{sec:coll}) and whether they would like to continue to use the solution for playing sheet music privately. Additional measures that not evaluated in this report but were measured include asking the participants how much they like the tool and whether they feel that they have learned the song.

Collected basic demographics were (age, gender), general interest in music (discrete 5-point Likert scale), prior experience with an instrument in years, prior experience with sheet music in years, and amount of instruments played. No other data points were collected.

\subsection{Data collection}
\label{sec:coll}
We used a repeated measurement design to maximize the amount of insight using our relatively small sample size.

Upon arrival at the lab, the participants were to complete a first form collecting data about the demographics. They would then be introduced to the experimental set-up and the task they were expected to accomplish (play along with the Flowpiano tool). Before the start of the experiment, they were able to pick a song tempo with which they felt comfortable. They then completed the first experiment mode (adaptive or base, decided by the flip of a coin) and completed an after-experiment form collecting the measures about flow, satisfaction with the tool, and whether they feel that they have learned the song. They then play the song with the tool, set to the other mode, and complete the same questionnaire as after the first time they have played the song. The participant then concludes the experiment by filling out a closure form measuring whether they would like to continue using the tool and which mode they have liked better.

\subsection{Quality of measures}
When playing the song, the participants were situated in a sound-proof room without the test supervisors present such that they do not feel observed while playing, thus removing this possible obstacle to attaining a flow state.
The questionnaire answers were pseudonymized, and the supervisors did not watch over the participants while they were completing the questionnaires but were available for questions. The authors supervised all of the tests.

%%%%%%%%%%%%%%%%%%%%%%%%%%%%%%%%%%%%%%%%%%%%%%%%%%%%%%%%%%%%%%%%%%%%%%%%%%%%%%%%
\section{RESULTS}
\begin{figure}[t]
\centering
\includegraphics[width=0.9\columnwidth]{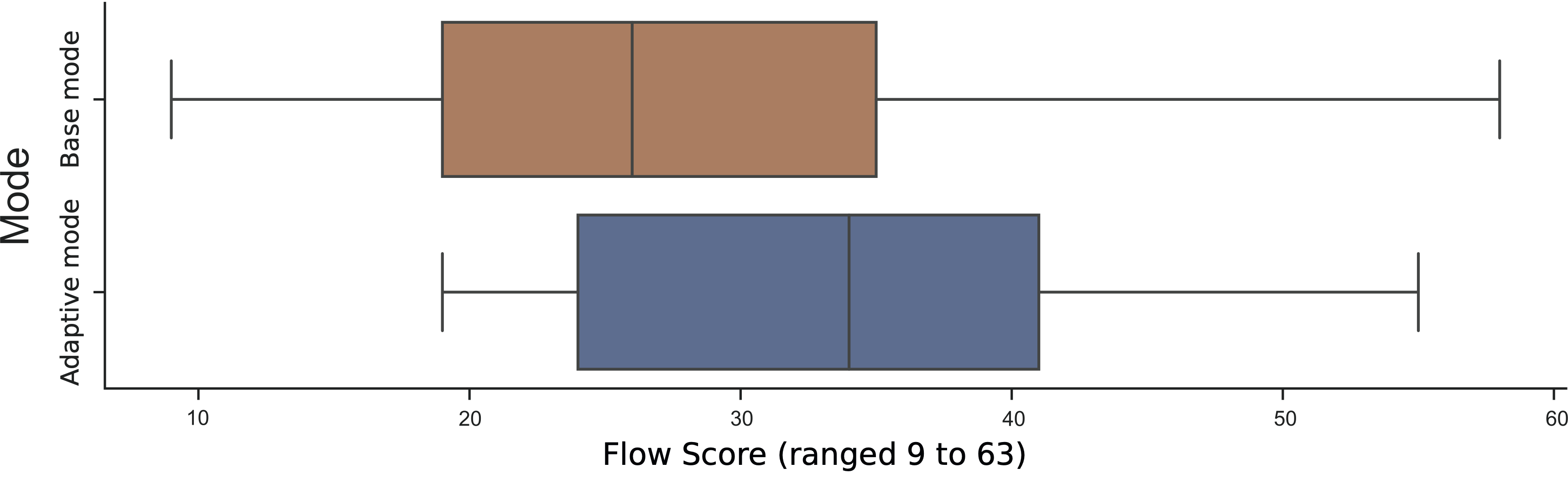}
\caption{Box-plot of Flow of participants according to used mode (adaptive vs. base mode). Lower and upper box boundaries 25th and 75th percentiles, respectively, line inside box median, lower and upper error lines 10th and 90th percentiles, respectively.}
\label{fig:dist}
\vspace{-2em}
\end{figure}

We can assert normality for the flow scores for both the adaptive and the base mode. Using a repeated-measures ANOVA test, one can also say that the underlying distributions of the flow scores for both modes are distinct from one another ($F(1, 20) = 5.932$, $p = .024$). The mean of the flow scores is at $\mu_f = 32.95$ (with $\sigma^2_f = 107.4$) for the adaptive mode and at $\mu_b = 27.95$ (with $\sigma^2_b = 172.4$; see Figure \ref{fig:dist}) for base mode.

We did not find any indication that the order of the modes had any impact on the flow scores. The median of years of experience on the piano amongst the participants who knew playing the instrument at all was at $\eta_\text{piano} = 3$. Those who can read sheet music were already doing that for a median of $\eta_\text{sheet} = 8$ years, and those who played an instrument practiced an instrument for a median of $\eta_\text{instr} = 8$ years. For each of those metrics, we split our participants into three cohorts: Those with no experience, less experience than the median number of years, or equal or more experience with the median.
We then tested for correlation (Pearson) between each of those group memberships with the flow scores in both modes. See Table \ref{tab:corr-musicalprop} for the results.

\begin{table}[t]
  \centering
  \caption{Pearson correlation test between group affiliation with respect to musical propensity and flow scores in the different experiment modes (\textit{control} and \textit{adaptive})}
  \resizebox{\columnwidth}{!}{
    \begin{tabular}{l|ll|ll}
    \toprule
\multicolumn{1}{l}{Group set} & \multicolumn{1}{l}{Control: r} & \multicolumn{1}{c}{Control: $p$} & \multicolumn{1}{l}{Adaptive: r} & \multicolumn{1}{c}{Adaptive: $p$}\\
    \midrule
    Piano practice     & $0.968$ & $ <0.001$ & $0.984$ & $ <0.001$\\
    Sheet music        & $0.666$ & $0.02$    & $0.994$ & $ <0.001$\\
    Instrument practice & $0.544$ & $0.04$    & $0.992$ & $ <0.001$\\
    \bottomrule
    \end{tabular}%
    }
  \label{tab:corr-musicalprop}%
\end{table}

 \begin{figure}
\subfloat[Piano experience]{\includegraphics[width=0.48\columnwidth]{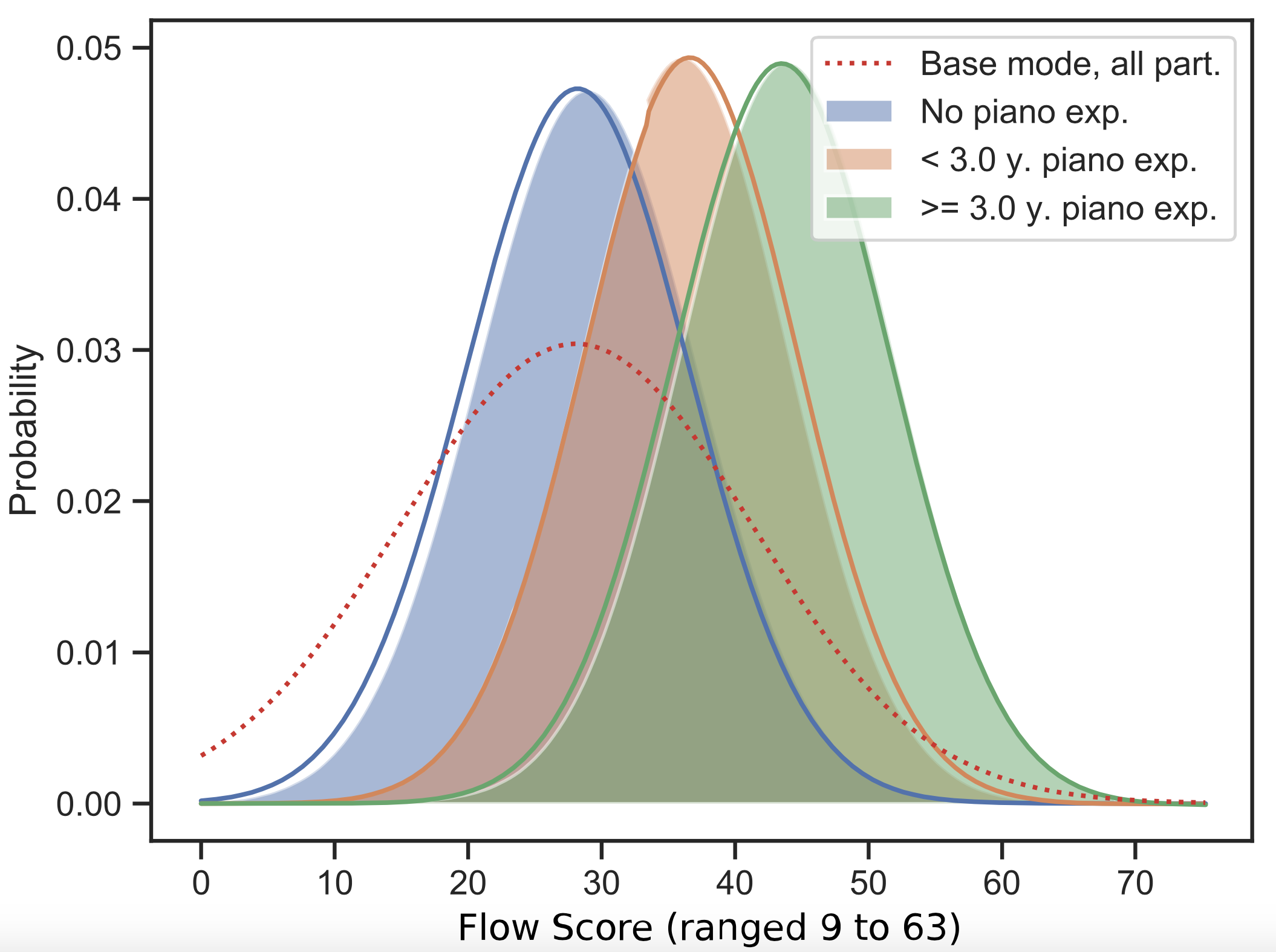}} 
%\label{fig:piano-groups}
\subfloat[Sheet music experience]{\includegraphics[width=0.48\columnwidth]{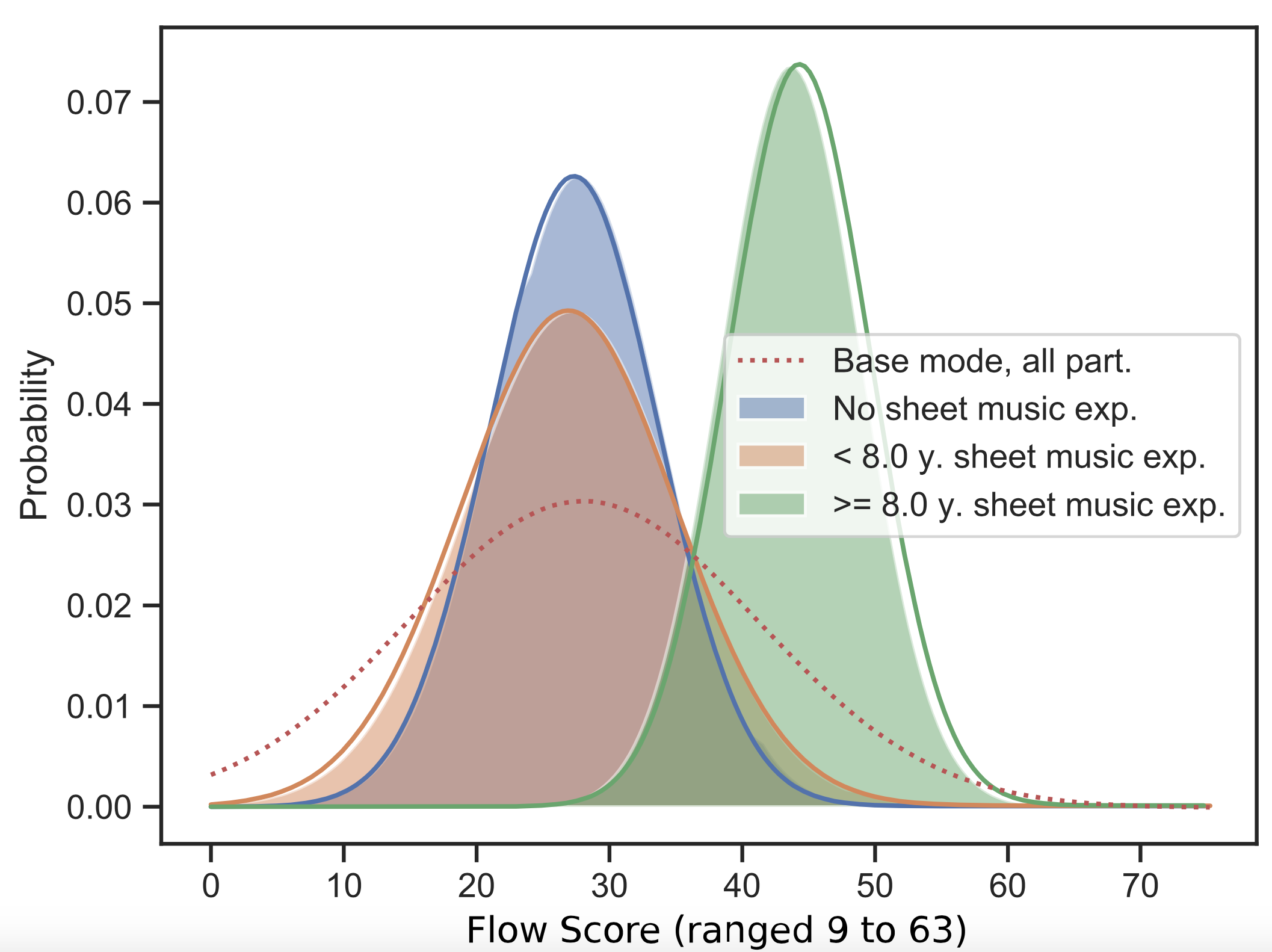}}
%\label{fig:sm-groups}
\caption{Flow score distributions for various musical skill groups}
\label{some example}
\end{figure}

Switching to the adaptive mode increases the strength of the correlation between the degree of musical propensity with respect to the group discriminator and the degree of flow experienced. 

This data can be visualized by looking at the distinct flow distributions for the piano groups ($p$). All of them are normal distributions with different means and variances (cf. Figure 3). The figure clearly shows that when the adaptive mode is enabled, more experienced pianists enter further into the flow state. A similar result can be obtained by looking at the sheet music groups ($s$). An ANOVA test shows that these differences are significant ($F_p(2, 18) = 6.448$, $p_p = .008$ and $F_s(2, 18) = 16.333$, $p_s < .001$)

Nine of the test participants (43\%) said that they wanted to use the solution in the future to practice sheet music reading and playing the piano, eleven said they would consider it (52\%, ``maybe'') and one participant ruled out using the solution in the future. All but one participant (96\%) said that they preferred the adaptive over the base mode. We were not able to show a correlation between high flow scores or musical propensity and wanting to continue to use the solution.

If not otherwise specified, all tests in this paper were conducted at a $p$-value of $0.05$.

%%%%%%%%%%%%%%%%%%%%%%%%%%%%%%%%%%%%%%%%%%%%%%%%%%%%%%%%%%%%%%%%%%%%%%%%%%%%%%%%
\section{DISCUSSION}
As we have shown, our participants experienced higher levels of flow while using the adaptive mode. This seems to be consistent with the results of Harmat et al. \cite{Harmat2015}.

While switching to adaptive mode seemed to increase the flow score across all levels of experience, for less experienced participants might be due to starting on a lower difficulty level, which more closely matches their skills. For more experienced participants, the higher levels of flow experienced might indicate that they were eased into playing the song by starting in adaptive mode. 

However, having less sheet music reading experience than the median seemed to provide no increase in the flow experience compared to participants with no sheet music reading experience. This might be due to the frustration participants experienced when seeing notes they used to be able to read but not being able to apply that skill swiftly enough anymore. Meanwhile, participants with no prior sheet music experience could have focused on the on-screen keyboard to play the notes, thus reducing frustration.

Unexpectedly, higher flow scores did not correlate with the wish for continued use of the solution. This seems counter-intuitive to the concept of flow, making activities more enjoyable. This might be due to higher flow scores correlation with music experience and thus presenting participants with a different form of display than what they practiced on.

%%%%%%%%%%%%%%%%%%%%%%%%%%%%%%%%%%%%%%%%%%%%%%%%%%%%%%%%%%%%%%%%%%%%%%%%%%%%%%%%
\section{CONCLUSION}
We demonstrated that visual feedback and difficulty scaling could positively impact the chance of musicians to experience flow while playing a new song of the piano. Musicians generally seem to prefer experiences with such feedback over those without it. It seems that design principles of other experiences like computer games with respect to flow can be transferred to musical tooling.
More experienced musicians were able to potentially profit more from the feedback that our solution provided than those who were more novice. A portion of our test participants -- but not the majority -- would like to continue to use such a solution at home. The amount of flow experienced had no impact on whether people want to continue practicing using a solution like ours.

In further research, we would like to obtain a set of participants with a more even distribution of musical experience to better adapt the song selection to them. Also, we might want to look at the relation between the difficulty metric we used to determine how difficult the voice played by the user should be and the reported flow. Additionally, it might be interesting to observe whether our approach helps with learning sheet music and learning to play a song.

\bibliographystyle{IEEEtran}
\bibliography{bibliography/bibliography.bib} 

\end{document}